\documentclass[iop]{emulateapj}

\slugcomment{}

\shorttitle{On the afterglow and progenitor of FRB 150418}
\shortauthors{Zhang}

\begin{document}


\title{On the afterglow and progenitor of FRB 150418}


\author{Bing Zhang}
\affil{Department of Physics and Astronomy, University of Nevada Las Vegas, NV 89154, USA}



\begin{abstract}
Keane et al. recently detected a fading radio source following FRB 150418, leading to the identification of a putative host galaxy at $z = 0.492 \pm 0.008$. Assuming that the fading source is the afterglow of FRB 150418, I model the afterglow and constrain the isotropic energy of the explosion to be a few $10^{50}$ erg, comparable to that of a short duration GRB. The outflow may have a jet opening angle of $\sim 0.22$ rad, so that the beaming-corrected energy is below $10^{49}$ erg. The results rule out most FRB progenitor models for this FRB, but may be consistent with either of the following two scenarios. The first scenario invokes a merger of an NS-NS binary, which produced an undetected short GRB and a supra-massive neutron star, which subsequently collapsed into a black hole, probably 100s of seconds after the short GRB. The second scenario invokes a merger of a compact star binary (BH-BH, NS-NS, or BH-NS) system whose pre-merger dynamical magnetospheric activities made the FRB, which is followed by an undetected short GRB-like transient. The gravitational wave (GW) event GW150914 would be a sister of FRB 150418 in this second scenario. In both cases, one expects an exciting prospect of GW/FRB/GRB associations.
\end{abstract}





\section{Introduction}

Fast radio bursts (FRBs) are high-Galactic-latitude radio bursting sources with anomalously high dispersion measure (DM) \citep{lorimer07,thornton13}. Due to the limited data, especially the lack of distance measurements, their physical origin has been unknown.

Lately, \cite{keane16} discovered a bright radio fading transient following FRB 150418, which lasted for about 6 days. This led to the identification of a putative host galaxy of the FRB\footnote{An objection to this claim was raised by \cite{williams16}, but see a counter-argument by \cite{li16}.}, which is an elliptical galaxy at $z=0.492\pm 0.008$. The cosmological origin of at least some FRBs, if confirmed by more observations, has profound implications. It opens the exciting prospects to use FRBs to probe the universe, including identifying missing baryons \citep{mcquinn14}, constraining the baryon number density $\Omega_b$ \citep{deng14a,keane16}, 
dark energy properties \citep{zhou14,gao14}, and the ionization history of the universe \citep{deng14a,zheng14}, as well as conducting fundamental tests of Einstein's Equivalent Principle \citep{wei15}, and setting a stringent upper limit on the photon mass \citep{wu16b}.

This {\em Letter} addresses another question: if the radio transient following FRB 150418 is indeed the afterglow of the FRB, what would the discovery tell us about the progenitor system of the FRB? I will use the observational data to constrain the energetics of the event, and subsequently narrow down the available progenitor models for 150418-like FRBs. 

\section{The afterglow}

According to \cite{keane16}, the brightness of the radio afterglow is comparable to that of short-duration gamma-ray bursts (GRBs). The follow-up observations were made using the Australia Telescope Compact Array (ATCA) at 5.5 GHz and 7.5 GHz. The first observation started at about 2 hours (7740 s) after the FRB and lasted for 19800 s. The mid-point epoch of this observation is at $t_1 = 17640$ s after the FRB. The flux density is $F_\nu (t_1) = 0.27 \pm 0.05$ mJy (per beam) at 5.5 GHz, and $F_\nu (t_1)  = 0.18 \pm 0.03$ mJy (per beam) at 7.5 GHz. The second observation started at almost 6 days later (512580 s after the FRB) and lasted for 72900 s. The mid-point epoch of this observation is $t_2 = 549030$ s. The 5.5 GHz flux density is $F_\nu (t_2) = 0.23 \pm 0.02$ mJy (per beam), and the source was not detected in 7.5 GHz, with an upper limit $F_\nu (t_2) < 0.08$ mJy (per beam). The third observation started about 2 more days later (681830 s after the FRB) and lasted for 74700 s. The mid-point epoch of this observation is at $t_3 = 719180$ s. The 5.5 GHz flux density is $F_\nu(t_3) \sim 0.09$ mJy (per beam), and the source was still not detected in 7.5 GHz. There were two more later observations, but the 5.5 GHz flux density is saturated around 0.09 mJy per beam, and 7.5 GHz observations only gave upper limits. One may regard 0.09 mJy per beam as the background flux from the host, then the 5.5GHz afterglow flux at $t_3$ should be below 0.09 mJy per beam. 

One may apply the 5.5 GHz data\footnote{The 7.5 Hz data are troublesome to the picture discussed below, and to the synchrotron theory in general. According to the data, the spectral slopes are $\beta_1 = 1.31^{+1.14}_{1.16}$ at $t_1$ and $\beta_2 > 3.40^{+0.27}_{-0.29}$ at $t_2$, with $\Delta \beta = \beta_2 - \beta_1 > 2.11^{+1.41}_{-1.44}$. This is much greater than $\Delta \beta = 0.5$ allowed by the synchrotron cooling theory. } to estimate the decay slope (in the convention of $F_\nu \propto t^{-\alpha}\nu^{-\beta}$) between the first two epochs: $\alpha_{21} = - \log(F_\nu(t_2)/F_\nu(t_1))/ \log(t_2/t_1) \sim 0.047^{+0.098}_{-0.090}$, with the error defined by the flux error and time bin size of each measurement. This is much shallower than the nominal decay slope ($\sim 1$) according to the standard GRB afterglow model \citep{meszarosrees97,sari98,gao13b}, suggesting that the peak time $t_p$ is likely between $t_1$ and $t_2$. Similarly, one can derive the decay slope between the second and third epoch, $\alpha_{32} = - \log(F_\nu(t_3)/F_\nu(t_2))/ \log(t_3/t_2) \geq 3.48^{+4.88}_{-1.81}$, with the $\geq$ sign reflecting the fact $F_\nu(t_3) \leq 0.09$ mJy. This decay slope is unusually steep for an external shock model, which may require a collimated jet \citep{rhoads99}. Alternatively, the source may undergo significant scintillation \citep[e.g.][]{williams16} during the second epoch of observation, so that the true decay slope may be shallower than observed.

Based on the standard afterglow model, the peak flux density of the afterglow is
\begin{equation}
F_{\rm \nu,max} = (0.51~{\rm mJy}) (1+z) \epsilon_{\rm B,-1}^{1/2} E_{\rm K,iso,51} n^{-1} D_{\rm L,28}^{-2},
\end{equation}
where $E_{\rm K,iso}$ is the isotropic kinetic energy of the outflow (normalized to $10^{51}~{\rm erg}$), $n$ is the ambient medium number density, $\epsilon_{\rm B}$ is the magnetic equipartition parameter (normalized to 0.1), and $D_{\rm L}$ is the luminosity distance. Plugging in $z=0.492$ and $D_{\rm L,28}=0.87$ (concordance cosmology adopted), and assuming $F_{\rm \nu,max} \gtrsim 0.27$ mJy, one can immediately derive 
\begin{equation}
E_{\rm K,iso} \gtrsim (2.7 \times 10^{50}~{\rm erg})~ n \epsilon_{\rm B.-1}^{-1/2}.
\label{eq:E}
\end{equation}
This is indeed an energy of the order of a short GRB. 

The peak time should correspond to the epoch when the minimum injection synchrotron frequency $\nu_m$ crosses the radio frequency 5.5 GHz. Expressing 
\begin{equation}
\nu_m \simeq (1.0\times 10^{11}~{\rm Hz})~ \left(\frac{1+z}{2}\right)^{1/2} \epsilon_{\rm B,-1}^{1/2} \epsilon_{\rm e,-1}^2 E_{\rm K,iso,51}^{1/2} t_5^{-3/2},
\end{equation}
where $\epsilon_e$ is the electron equipartition parameter (normalized to 0.1), and $t$ is the observer time (normalized to $10^5$ s). Requiring $\nu_m = 5.5$ GHz and making use of Eq.(\ref{eq:E}), one can derive the peak time
\begin{equation}
t_{\rm peak} \gtrsim (4.1\times 10^5~{\rm s})~ \epsilon_{\rm B,-1}^{1/6} \epsilon_{\rm e,-1}^{4/3} n^{1/3}.
\end{equation}
This is smaller than $t_2$, which is consistent with the data. 

Finally, if one assumes that the steep $\alpha_{23}$ is intrinsic (rather than a consequence of scintillation), one may constrain the jet opening angle $\theta_j$ by requiring the jet break time $t_j \lesssim t_2$, i.e.
\begin{equation}
\theta_j \simeq (0.084~{\rm rad})~ \left(\frac{t_j}{1~{\rm day}}\right)^{3/8} \left(\frac{1+z}{2}\right)^{-3/8} E_{\rm K,iso,51}^{-1/8} n^{1/8}.
\end{equation}
Applying Eq.(\ref{eq:E}) and $t_j \leq t_2$, one gets
\begin{equation}
\theta_j \lesssim 0.22 \epsilon_{\rm B,-1}^{1/16}.
\end{equation}
The beaming-corrected kinetic energy is
\begin{equation}
E_{\rm K} = E_{\rm K,iso} (1-\cos\theta_j) \sim (6.5 \times 10^{48}~{\rm erg})~n \epsilon_{\rm B,-1}^{-1/2}.
\label{eq:Ek}
\end{equation}

\section{Possible progenitor}

A possible association between FRBs and GRBs, especially short GRBs, was first proposed in \cite{zhang14}\footnote{\cite{totani13} first proposed a possible association between FRBs and NS-NS mergers, but most of his FRBs are not supposed to be associated with GRBs. }. It has been shown that the afterglow of FRBs, without association of a GRB-like event, would be quite faint \citep{yi14}. The association of FRB 150418 with a bright radio afterglow comparable to that of a short GRB therefore rules out most of the proposed progenitor systems for this FRB, such as stellar flares \citep{loeb14}, giant radio pulses from young pulsars \citep{cordes16,connor16}, magnetar giant flares \citep{popov13,kulkarni14,katz15}, merger of double white dwarf systems \citep{kashiyama13b}, and comets falling onto neutron stars \citep{geng15}\footnote{According to \cite{keane16} and \cite{spitler16}, there are likely more than one type of FRBs. These progenitor models may be still valid for other types of FRBs.}. The standard ``blitzar" model \citep{falcke14} has a maximum energy of the order of the total magnetic field energy in the ejected magnetosphere, which is $\sim (1.7\times 10^{47} {\rm erg})B_{p,15}^2 R_6^3$ \citep{zhang14}. It might reach the beaming-corrected energy of FRB 150418 if the surface polar cap magnetic energy can reach $B_p \sim 6.2 \times 10^{15}$ G. However, with a large delay time between the supra-massive neutron star birth and collapse, as envisaged by \cite{falcke14}, the environment of the neutron star would be very clean. It is hard to collimate the outflow to the desired small jet opening angle. Without collimation, the isotropic energy ($E_{\rm K,iso} \sim 2.7\times 10^{50}$ erg) of the FRB requires $B_{p} > 4 \times 10^{16}$ G to power the afterglow, which is far-stretching. On the other hand, for the scenario of a blitzar following a short GRB, as proposed by \cite{zhang14}, the energy associated with the pre-blitzar explosion would be large enough to power the afterglow, without introducing extreme parameters to the supra-massive neutron star itself (see more discussion below in Sect. 3.2).
In general, the large energy budget of the radio afterglow requires that the progenitor system involves a catastrophic event. The relatively small energy as compared with long GRBs, and more importantly, the elliptical host galaxy, point towards a compact star merger event. Below I discuss several possible candidates.

\subsection{NS-NS mergers with a post-merger BH and BH-NS mergers}

Traditional NS-NS mergers are expected to produce a black hole as the post-merger product. Accretion of tidally disrupted debris into the newly formed BH would power a short GRB \citep{paczynski86,eichler89,rezzolla11}. No FRB is expected. The same applies to BH-NS mergers \citep{paczynski91}.

For NS-NS mergers, it is possible that a differential-rotation-supported hyper-massive NS may survive for 100s of milliseconds \citep[e.g.][]{hotokezaka13}. It later collapses to a BH, and the subsequent accretion powers a short GRB. \cite{totani13} argued that during the hyper-massive NS phase, synchronization of magnetic fields of the two neutron stars may produce an FRB. He envisaged isotropic emission of the FRB to interpret the high event rate of FRBs. However, significant mass ejections are expected before merger \citep[e.g.][]{hotokezaka13}, so that in most solid angles, an FRB, if produced, cannot penetrate through the ejecta and hence, cannot be observed \citep{zhang14}. Even in the polar direction where the ejecta may not block the FRB emission, a strong neutrino-driven baryon wind is expected, so that the plasma density is again too high to allow the propagation of FRB emission. 

I conclude that this ``standard" scenario of NS-NS/BH-NS mergers cannot account for FRB 150418-like events.

\subsection{NS-NS mergers with a supra-massive NS merger product}

If the neutron star equation of state is stiff enough, at least some NS-NS mergers would produce rigid-rotation-supported supra-massive NSs or stable NSs \citep[e.g.][]{dai06,metzger08,metzger11,bucciantini12,zhang13,rezzolla15,ciolfi15}. As a supra-massive NS spins down via either magnetic dipole radiation or gravitational wave radiation due to a magnetic-field-induced ellipticity \citep[e.g.][]{fan13,gao16,lasky16}, the internal dissipation within the pulsar wind might drive bright X-ray emission via synchrotron radiation (e.g. \citealt{zhang13} and references therein). After losing significant angular momentum, it would collapse to a BH at a critical rotation period, making an extremely steep decay segment in the X-ray afterglow light curve of a short GRB \citep{rowlinson13,lvzhang14,lv15}. Following the blitzar idea \citep{falcke14}, \cite{zhang14} suggested that an FRB may be launched as the supra-massive NS collapses and the magnetosphere ejected. This idea was further developed by \cite{ravi14}. The collapse time and the X-ray plateau luminosity can be used to constrain NS equation of state and the properties of the merger product \citep{lasky14,lv15,gao16}. 

Such a scenario is allowed for FRB 150418. The initial merger leads to a short GRB, and a later collapse leads to an FRB. The short GRB may not be bright enough to trigger GRB detectors (e.g. Fermi/GBM). Given $E_{\rm K,iso} \simeq 2.7\times 10^{50}$ erg, the isotropic $\gamma$-ray energy ($E_{\rm \gamma,iso}$) can be smaller than $10^{50}$ erg if the $\gamma$-ray emission efficiency is not large (e.g. $<10\%$), so that a short GRB may be too faint to be detected by Fermi/GBM.

One caveat of this scenario is the small energetics of the event. Since the post-merger supra-massive NS must be rapidly spinning, in order to account for the observed small energetics, the supra-massive NS must collapse before most spin energy is released, and/or most spin energy must be carried away through gravitational wave radiation \citep{gao16}.

\subsection{Pre-merger dynamical magnetospheric activities}

The possible detection of a 1 s - duration hard transient \citep{connaughton16} 0.4 s after the BH-BH merger gravitational event GW 150914 \citep{GW150914} raised a wave of modeling possible electromagnetic counterparts from BH-BH mergers. One scenario as suggested by \cite{zhang16} is that if at least one of the two BHs carries a certain charge,  the merging BHs would form a current loop, and thereby drive a magnetosphere with increasing wind power right before the final merger. The wind may power a putative short-duration GRB if the charge is large enough (with a dimensionless charge $\hat q \sim 10^{-4}$), or an FRB for a smaller threshold charge ($\hat q > 10^{-7}$). This model predicts a direct association of FRBs with short GRBs along with GW events. A related mechanism invoking charged Kerr-Newman black holes was proposed by \cite{liu16}. The spirit of \cite{zhang16} model, namely, launching a Poynting flux from a dynamically evolving electromagnetic system due to a merger, is also applicable to other scenarios that invoke magnetospheres rather than charges. For example, \cite{wang16} recently proposed an FRB model invoking two merging NSs. They suggested that the electromotive force induced due to the inspiral of the system would accelerate electrons to power an FRB. In general, these models invoking pre-merger dynamical magnetospheric activities present another plausible type of scenarios to interpret FRB 150418-like events.

Observationally, the isotropic kinetic energy of FRB 150418 (Eq.(\ref{eq:E})) is about one order of magnitude larger than that of the EM energy of GW/GBM 150914 event \citep{connaughton16}, but the beaming-corrected kinetic energy (Eq.(\ref{eq:Ek})) is slightly smaller. The similarity of energetics between the two events suggests a possible connection between GW 150914-like events and FRB 150418-like events.

The steep decay slope $\alpha_{23}$, if not an effect of scintillation, would be difficult to interpret with a BH-BH merger, since the environment is relatively clean and collimation of the jet is not easy. Strong collimation is possible for a NS-NS or NS-BH merger system, since the ejecta launched during the inspiral phase would serve as a collimator of the jet.

\section{Event rate density}

The current FRB event rate density may be estimated as
\begin{eqnarray}
\dot \rho_{\rm FRB} & = & \frac{365 f \dot N_{\rm FRB}}{(4\pi/3) D_{\rm L}^3} \nonumber \\
& \simeq & (720 {\rm Gpc^{-3}~yr^{-1}}) f \left(\frac{D_{\rm L}}{6.7~{\rm Gpc}}\right)^{-3} \left(\frac{\dot N_{\rm FRB}}{2500}\right),
\label{eq:rho-FRB}
\end{eqnarray}
where $\dot N_{\rm FRB}$ is the daily all-sky FRB rate which is normalized to 2500 \citep{keane15}, and $D_{\rm L}$ is the luminosity distance of the FRB normalized to 6.7 Gpc ($z=1$). Since there are more than one types of progenitor \citep{keane16,spitler16}, the parameter $f<1$ denotes the fraction of 150418-like FRBs. 

For the NS-NS merger scenario leading to a supra-massive NS which produces an FRB as it collapses \citep{zhang14}, one may expect that the rate of observable short GRB/FRB associations is comparable to the observed short GRB rate, $(4-7)~{\rm Gpc^{-3}yr^{-1}}$ \citep{sun15}, corrected for the fraction that produces a supra-massive NS (about 30\%, \citealt{gao16}). This falls short to the observed value in Eq.(\ref{eq:rho-FRB}). However, since the putative short GRB associated with FRB 150418 did not trigger GRB detectors, there could be a large population of un-triggered GRB-like events that may be associated with FRBs. If this blitzar mechanism is correct, there should be also other FRBs that are generated from supra-massive NSs with a much larger delay time from birth to death \citep{falcke14}. Since the total energy budget of these blitzars is much smaller than that of FRB 150418, their radio afterglows would be much fainter than this one \citep{yi14}. The first detection of FRB 150418 radio afterglow may be simply due to a selection effect.

Within the pre-merger dynamical magnetospheric activity scenarios \citep{zhang16,liu16,wang16}, the FRB event rate is directly related to the compact star merger event rate. 
The event rate density of BH-BH mergers as inferred by the LIGO team based on the detection of GW 150914 is $\sim (2-53) ~{\rm Gpc}^{-3}~{\rm yr}^{-1}$, with the most optimistic value as high as $\sim 400 ~{\rm Gpc}^{-3}~{\rm yr}^{-1}$ \citep{GW-rate}. This is somewhat smaller than Eq.(\ref{eq:rho-FRB}), but could become consistent with the observed rate if the cosmological fraction $f$ is small enough. More comfortably, the theoretical event rate can be boosted up by a factor of a few if one includes NS-NS and NS-BH mergers into the mix. The true rates of these two types of mergers are at least one order of magnitude higher than BH-BH mergers. However, considering possible beaming in these systems due to the ejecta surrounding the system, the net gain of the FRB event rate density by including these systems may be a factor of a few.

\section{Summary and discussion}

Assuming that the radio transient following FRB 150418 is its afterglow, I have modeled the radio afterglow data and reached the following conclusions: The afterglow demands a large isotropic kinetic energy of several $\times 10^{50}$ erg. If the steep decay after $t_2 \simeq 5.5 \times 10^5$ s is interpreted as due to a jet break, the inferred jet opening angle is $\lesssim 0.22$ rad, and the beaming corrected energy drops below $10^{49}$ erg.

In view of the energetics and the observed elliptical host galaxy, the progenitor system of FRB 150418 should invoke a compact star merger. The traditional model of generating GRBs through accretion into a BH merger product is not a credential candidate to produce FRBs (c.f. \citealt{totani13}). The data, on the other hand, seem to be consistent with either of the following two scenarios: 1. An NS-NS merger leads to a supra-massive NS after the merger, which subsequently collapses into a BH and releases an FRB \citep{zhang14}. The expectations of this scenario include that the FRB should lag behind the GRB by some time (typically hundreds of seconds based on observations, \citealt{gao16} and references therein), and that there should be other blitzar-like events that are not associated with GRBs \citep{falcke14}. 2. The FRB is produced before the merger of two compact objects through some pre-merger dynamical magnetospheric activities \citep{zhang16,liu16,wang16}. Within this scenario, the FRB may slightly lead the short GRB, but may coincide with the GW chirp signal. The LIGO event GW 150914 \citep{GW150914} would be a sister of FRB 150418. One would expect an FRB simultaneously emitted at the GW chirp signal time of GW 150914, and a GW chirp signal at the time of FRB 150418. 

In both cases, one would foresee an exciting prospect of GW/FRB/GRB associations. A joint search among the GW, FRB, and GRB communities is encouraged. The joint detections/non-detections would prove/disprove the models discussed here, and the relative ordering among the GW, FRB, and GRB signals (for detections) would differentiate the different scenarios discussed in this {\em Letter}.

Finally, all the discussion in this {\em Letter} is based on the assumption that the fading radio transient following FRB 150418 is the afterglow of the FRB. \cite{williams16} argued that the putative radio afterglow of FRB 150418 is simply emission of a flaring AGN that happens to fall into the Parkes beam of FRB 150418. On the other hand, \cite{li16} showed that the chances to have an AGN with such a high variability in the Parkes beam of FRB 150418 and to have a bright flare right after the FRB are low. They argued that the afterglow possibility is not ruled out. Further monitoring of the source and follow-up observations of other FRBs can tell whether some FRBs (FRB 150418-like) are indeed followed by bright radio afterglows, and hence, whether the discussion in this {\em Letter} is relevant.




\acknowledgments
I thank Matthew Bailes, Edo Berger, Jonathan Katz, Even Keane, Ye Li, and Z. Lucas Uhm for helpful discussion/comments, and the anonymous referee for helpful suggestions.
This work is partially supported by NASA NNX15AK85G and NNX14AF85G.

\end{document}